\documentclass[]{aa}
\usepackage{graphicx}
\usepackage{natbib}

\bibpunct{(}{)}{;}{a}{}{,}

\begin{document}

   \title{The ISO Long Wavelength Spectrometer line spectrum of VY Canis Majoris and other oxygen-rich evolved stars \thanks{Based on observations with ISO, an ESA project with instruments funded by ESA Member States (especially  the PI countries: France, Germany, the Netherlands and the United Kingdom) with the participation of ISAS and NASA.}
   }

   \author{E. T. Polehampton
           \inst{1,2,3}
           \thanks{Current address: Space Science Department, Rutherford Appleton Laboratory, UK}
           \and
           K. M. Menten
           \inst{1}
           \and
           F. F. S. van der Tak
           \inst{4, 1}
          \and
           G. J. White
           \inst{2,5}
          }
          \institute{
            Max-Planck-Institut f\"{u}r Radioastronomie, Auf dem H\"{u}gel 69, 53121 Bonn, Germany 
            \and
            Space Science \& Technology Department, Rutherford Appleton Laboratory, Chilton, Didcot, Oxfordshire, OX11 0QX, UK
            \email{edward.polehampton@stfc.ac.uk}
            \and
            Department of Physics, University of Lethbridge, 4401 University Drive, Lethbridge, Alberta, T1J 1B1, Canada 
            \and
            Netherlands Institute for Space Research (SRON), Landleven 12, 9747 AD Groningen, The Netherlands 
            \and 
            Department of Physics and Astronomy, The Open University, Milton Keynes, MK6 7AA, England.
          }

   \date{Received  / accepted }

 \abstract{The far-infrared spectra of circumstellar envelopes around various oxygen-rich stars were observed using the ISO Long Wavelength Spectrometer (LWS). These have been shown to be spectrally rich, particularly in water lines, indicating a high H$_{2}$O abundance.}
{We have examined high signal-to-noise ISO LWS observations of the luminous supergiant star, VY CMa, with the aim of identifying all of the spectral lines. By paying particular attention to water lines, we aim to separate the lines due to other species, in particular, to prepare for forthcoming observations that will cover the same spectral range using Herschel PACS and at higher spectral resolution using Herschel HIFI and SOFIA.}
{We have developed a fitting method to account for blended water lines using a simple weighting scheme to distribute the flux. We have used this fit to separate lines due to other species which cannot be assigned to water. We have applied this approach to several other stars which we compare with VY CMa}
{We present line fluxes for the unblended H$_2$O and CO lines, and present detections of several possible $\nu_{2}=1$ vibrationally excited water lines. We also identify blended lines of OH, one unblended and several blended lines of NH$_{3}$, and one possible detection of H$_{3}$O$^{+}$.}
{The spectrum of VY CMa shows a detection of emission from virtually every water line up to 2000 K above the ground state, as well as many additional higher energy and some vibrationally excited lines. A simple rotation diagram analysis shows large scatter (probably due to some optically thick lines). The fit gives a rotational temperature of 670$^{+210}_{-130}$~K, and lower limit on the water column density of $(7.0\pm1.2)\times10^{19}$~cm$^{-2}$. We estimate a CO column density $\sim$100 times lower, showing that water is the dominant oxygen carrier. The other stars that we examined have similar rotation temperatures, but their H$_2$O column densities are an order of magnitude lower (as are the mass loss rates).}


\titlerunning{The ISO LWS line spectrum of VY CMa}
\authorrunning{E. T. Polehampton et al.}
\maketitle
%
%

\section{Introduction}

Circumstellar envelopes form around red giant and supergiant stars during the later stages of their evolution. The mass loss from the stars results in an expanding envelope of material consisting of dust and molecules whose absorption and/or emission can be observed from infrared through radio wavelengths. In oxygen-rich stars with high mass loss rates, the cooling of the envelope is believed to be dominated by rotational lines from water (H$_2$O) vapour \citep[e.g.][]{justtanont94}, which has a rich far infrared (FIR) spectrum. The FIR range also contains the rotational transitions of other simple molecules and molecular ions.

The Infrared Space Observatory \citep[ISO;][]{kessler} satellite has revolutionised the study of circumstellar envelopes by allowing access to all the spectral features in the mid- and far-infrared range. ISO spectra observed using the Short Wavelength Spectrometer \citep[SWS;][]{degraaw} and Long Wavelength Spectrometer \citep[LWS;][]{clegg} have previously been analysed, showing that the majority of the detected features in the FIR range are due to the rotational transitions of water (W Hya SWS: \citet{neufeld_d}; W Hya LWS: \citet{barlow}; R Cas LWS: \citet{truongbach}; VY CMa SWS: \citet{neufeld_e}).

\begin{table}
\caption{Characteristics of the stars used in this paper.}
\label{charcs}
\begin{tabular}{lccc}
\hline
\hline
Star name          & Distance & Mass Loss Rate  & Ref.  \\ 
                   &  (pc)    & (M$_{\odot}$ yr$^{-1}$)  & \\
\hline
VY CMa             & 1100     & 0.5--1$\times10^{-4}$         & (1), (2)\\
IK Tau             & 300      & 4.9$\times10^{-6}$            & (3)\\     
TX Cam             & 390      & 2.6$\times10^{-6}$            & (3)\\
RX Boo             & 160      & 2.2$\times10^{-7}$            & (4)\\
IRC+10011 (WX Psc) & 830      & 6$\times10^{-6}$              & (5)\\
R Cas              & 107      & 3.4$\times10^{-7}$            & (6)\\
\hline
\end{tabular}

\medskip
(1) Reid \& Menten (in preparation); \citet{choi}; (2) \citet{decin}; (3) \citet{olivier}; (4) \citet{teyssierAgb}; (5) \citet{decinWXPsc}; (6) \citet{truongbach}.
\end{table}

In this paper, we present a dataset containing the ISO LWS spectra of the red supergiant star VY CMa, and several other evolved stars. The basic parameters of these spectra, such as identifications of lines over the wavelength range 45--$196~\mu$m and their fluxes are discussed here. We concentrate on the results from VY CMa, as it shows the spectrum with the highest signal to noise ratio and the line identifications are typical of several other stars in the sample. The SWS line spectrum for this star has previously been presented by
\citet{neufeld_e}. The continuum emission of the SWS and LWS spectra has been modeled by \citet{harwit_b} and the LWS line emission has
been briefly described by \citet{barlow_b}. VY CMa is thought to be in a late stage of stellar evolution, and appears extremely bright at
infrared wavelengths. It has a mass loss rate estimated to be of the order of $5\times10^{-5}$ to $1\times10^{-4}$~M$_\odot$~yr$^{-1}$ with a period of higher mass loss in the past \citep{decin}. It is located at the edge of a molecular cloud \citep{lada}, and has a distance derived from H$_2$O and SiO masers of $\sim$1.1~kpc from the Sun \citep[][Reid \& Menten, in preparation]{choi}. In Sect.~\ref{vycma}, we estimate the column densities of water and CO in the envelope of VY CMa to determine which is the dominant oxygen carrier. In Sect.~\ref{otherstars}, we compare the results from VY CMa with several other stars that were also observed with the ISO LWS. The characteristics of these stars are shown in Table~\ref{charcs}.

\begin{table*}
\caption{\label{obsns} Log of the L01 observations used in this paper.}
\begin{center}
\begin{tabular}{lcccc}
\hline
\hline
Star            &  ISO TDT & Integration Time (s) & Date   & Coordinates (J2000) \\
\hline
VY CMa          & 73502338 & 1930 & 20-11-97 & 07$^{\rm{h}}$22$^{\rm{m}}$58.21$^{\rm{s}}$, -25$^{\circ}$46$^{'}$02.6$^{''}$ \\
IK Tau          & 65601707 & 3430 & 02-09-97 & 03$^{\rm{h}}$53$^{\rm{m}}$28.86$^{\rm{s}}$,  11$^{\circ}$24$^{'}$22.3$^{''}$ \\
TX Cam          & 69501069 & 3428 & 10-10-97 & 05$^{\rm{h}}$00$^{\rm{m}}$50.37$^{\rm{s}}$,  56$^{\circ}$10$^{'}$52.5$^{''}$\\
RX Boo          & 62300514 & 1328 & 31-07-97 & 14$^{\rm{h}}$24$^{\rm{m}}$11.62$^{\rm{s}}$,  25$^{\circ}$42$^{'}$13.9$^{''}$  \\
                & 59701307 & 1630 & 05-07-97 & 14$^{\rm{h}}$24$^{\rm{m}}$11.62$^{\rm{s}}$,  25$^{\circ}$42$^{'}$13.9$^{''}$ \\
IRC+10011       & 57701103 & 2796 & 15-06-97 & 01$^{\rm{h}}$06$^{\rm{m}}$25.96$^{\rm{s}}$,  12$^{\circ}$35$^{'}$53.2$^{''}$ \\
                & 57700513 & 1910 & 15-06-97 & 01$^{\rm{h}}$06$^{\rm{m}}$25.96$^{\rm{s}}$,  12$^{\circ}$35$^{'}$53.2$^{''}$\\
R Cas           & 56801440 & 2204 & 06-06-97 & 23$^{\rm{h}}$58$^{\rm{m}}$24.38$^{\rm{s}}$,  51$^{\circ}$23$^{'}$18.1$^{''}$ \\
\hline
\end{tabular}
\end{center}
\end{table*}

\begin{figure*}
\resizebox{18.8cm}{!}{\includegraphics{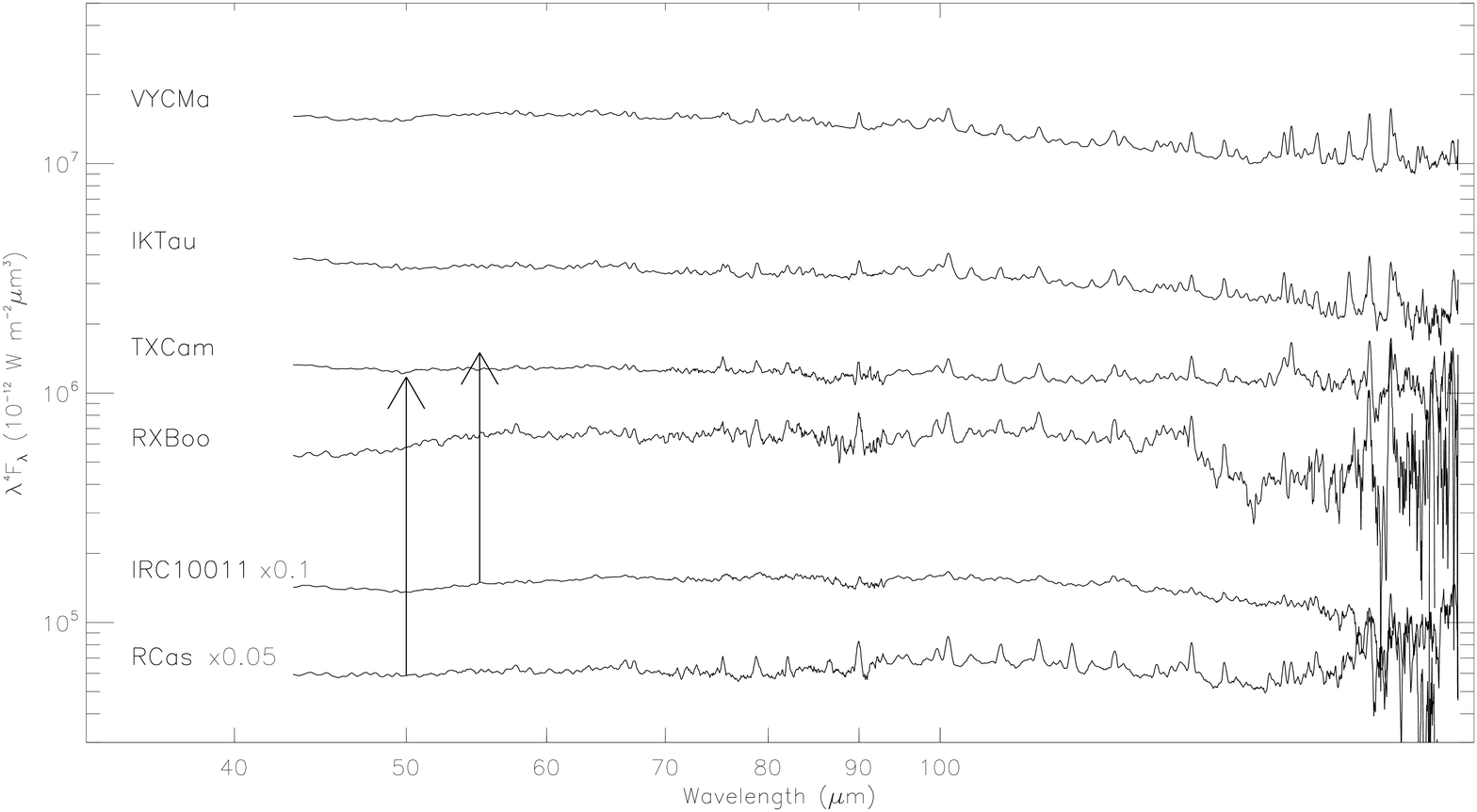}}
\caption{LWS spectra (45--197~$\mu$m) of the stars used in this paper in units of $\lambda^4f_{\lambda}$. From top to bottom: VY CMa, IK Tau, TX Cam, RX Boo, IRC+10011 ($\times$0.1) and R Cas ($\times$0.05). The spectra of IRC+10011 and R Cas have been multiplied by factors of 0.1 and 0.05, respectively for clarity in the plot. The arrows indicate the true flux level for these stars before the multiplication factor was applied. Each spectrum was smoothed to reduce the noise before plotting.}
\label{contm}
\end{figure*}

\section{Observations and data reduction \label{obs}}

A complete set of uniformly processed LWS grating spectra (using ISO observing mode L01) have been uploaded to the ISO Data Archive\footnote{see {\tt www.iso.esac.esa.int/ida/}} as highly processed data products \citep[HPDP;][]{lloyd_d}. This allows us to make a detailed comparison of consistently and accurately calibrated spectral lines observed toward a sample of different stars. The objects, positions and ISO target dedicated time (TDT) numbers of the data we have used are shown in Table~\ref{obsns}.

To check the accuracy of the standard HPDP pipeline processing for our individual objects, we downloaded and manually reduced the LWS L01 spectrum of VY CMa, comparing it with the HPDP data (to examine the effects of automatically removing glitches and dark current subtraction). Both the test observations and the HPDP were initially processed with the LWS off line pipeline (OLP) version 10. 

In order to reduce the data by hand, we used the LWS Interactive Analysis \citep[LIA;][]{lim_d} software to optimise the dark current subtraction for each LWS detector. The absolute responsivity correction was not adjusted interactively because the OLP data already contains a better method of calculating the corrections to that available in the LIA software. Glitches were carefully removed from each scan by hand using the ISO Spectral Analysis Package \citep[ISAP;][]{sturm}. In addition, forward and backward scans were compared and the regions where one direction was affected by the response time of the detectors were removed (this can occur when the grating was scanned in the direction of increasing detector response from the edge of each band). A correction for fringing was applied using the LWS defringing algorithm within ISAP. The forward and reverse scans for each detector were then averaged.

A careful comparison of the interactively reduced and HPDP spectra showed that there were only minor differences in the final strength of the lines. This is mainly due to the defringing step which was only performed in the HPDP processing for detectors that the pipeline judged significantly affected (in the case of VY CMa, only for detector SW5, whereas interactively we judged defringing to make most difference to detector LW4). Further differences that may cause problems are that in the HPDP data reduction, both scan directions were simply averaged together (without accounting for the transient response of the detectors) and deglitching was carried out automatically. However, neither of these appear to have led to the introduction of spurious features in the HPDP spectrum.

The largest difference between the two data reduction techniques is an improvement in the spectral shape of data from detector SW1 in the HPDP data. This is due to an additional correction in the HPDP pipeline that removes a double-peaked structure (with peaks at $\sim$45~$\mu$m and $\sim$48~$\mu$m) that is believed to be due to spurious features in the SW1 relative spectral response function \citep{lloyd_d}. Using this correction allows us to have much greater confidence in the spectral shape observed using SW1 between 43 and 50~$\mu$m.

In conclusion, we determined that the HPDP pipeline did not introduce spurious artifacts that could be mistaken for lines, and that it had superior calibration for detector SW1. Therefore, we have used only the HPDP data for all of the stars in our sample. We applied multiplicative shifts to individual detector data to align their continua by examining the overlap region between pairs of detectors. However, this was not important for the spectrum of VY CMa because the absolute flux calibration applied by the HPDP already gave very good agreement between adjacent detectors, with the only exception being SW1, which had an absolute flux $\sim$20\% higher than SW2. For the other stars, the shifts were more often required for the long wavelength detectors (up to $\sim$30\%). In order to examine the line fluxes, we subtracted the continuum level by fitting a 3rd order polynomial baseline to the data from each detector independently.

Several well known spurious features that resemble lines remain in the spectra, and these are labeled in the plots presented here (e.g. features in absorption at 77~$\mu$m using SW5 and 191~$\mu$m using LW5, and features in emission at 107~$\mu$m and 109~$\mu$m using LW2; T. Grundy, private communication). Some of these lines occur in the overlap region between two detectors and so can be ignored by selecting the optimum wavelength ranges to use. To provide a further check that the line features we have identified are real, we compared the spectra to observations of the asteroid Ceres, which should not contain any spectral line features (but does contain the spurious features). 

Where there were multiple LWS observations of the sample stars that showed good agreement, these were co-added to increase the signal to
noise ratio. This applied to IRC+10011 and RX Boo (see Table~\ref{obsns}). Although several other observations were also available for R Cas (TDTs 26300712, 56801715 and 56801552), we have used only the one with longest integration time (TDT 56801440).

The LWS beam size was $\sim$80$^{''}$, and the spectral resolution was 0.29~$\mu$m for detectors SW1--SW5 and 0.6~$\mu$m for detectors LW1--LW5 \citep{gry}. The RMS noise in the spectra varies for the different stars, but is roughly (10--20)$\times10^{-20}$~W~cm$^{-2}$~$\mu$m$^{-1}$ in the short wavelength detectors (measured between 87 and 89.5~$\mu$m), and (0.5--2.2)$\times10^{-20}$~W~cm$^{-2}$~$\mu$m$^{-1}$ in the long wavelength detectors (measured between 140 and 142~$\mu$m). The wavelength bin width is 0.04~$\mu$m for the short wavelength detectors and 0.13~$\mu$m for the long wavelength detectors. The final spectra before continuum subtraction are shown in Fig.~\ref{contm}, plotted as $\lambda^4f_{\lambda}$. In this figure, the spectra have been smoothed with a window of 5 wavelength bins.

\section{Results: VY CMa \label{vycma}}

In this paper we concentrate on identifying and characterising the spectral lines in the LWS data between 45 and 196~$\mu$m. The following sections present the line identifications for VY CMa in this range.

\begin{figure*}
\begin{center}
\resizebox{!}{19.6cm}{\includegraphics[angle=0]{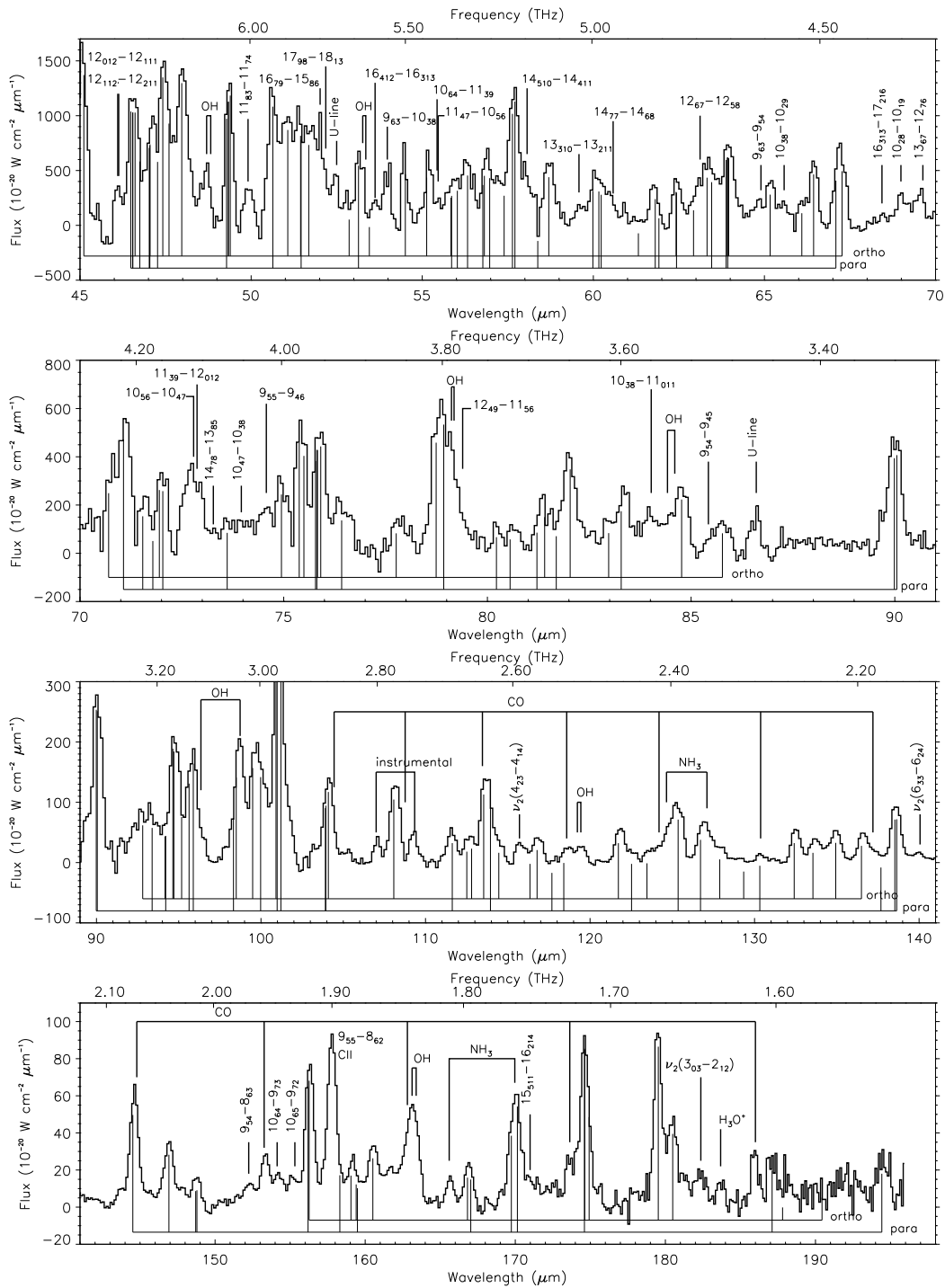}}
\end{center}
\caption{\label{vycmallines}Continuum subtracted LWS spectrum of VY CMa with all water line transitions with upper levels up to 2000 K above the ground state marked {\it below} the spectrum (para: lower, ortho: upper of the two groups) and other lines marked individually {\it above} the spectrum.}
\end{figure*}

\begin{figure*}
\begin{center}
\resizebox{15.0cm}{!}{\includegraphics[angle=0]{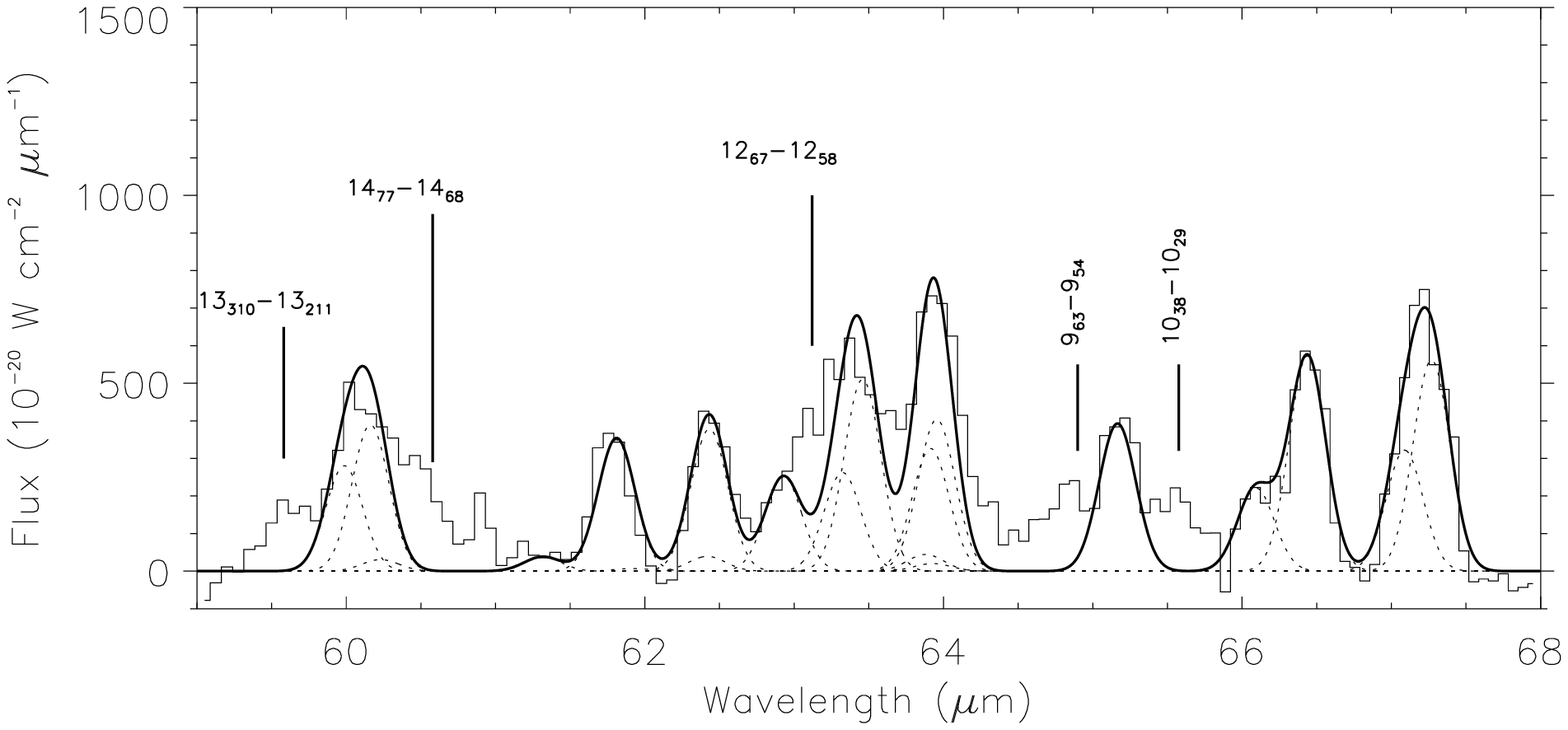}}
\end{center}
\caption{\label{show_fit}An example of our water line fit for VY CMa for the region 59--68~$\mu$m. The final fit is shown by the solid black line, and individual water lines by the dashed line. Additional transitions above 2000~K in energy are marked.}
\end{figure*}

\subsection{Water}

The SWS spectrum of VY CMa has already been shown to be rich in the rotational lines of water \citep{neufeld_e}. They identified at least 41 features due to water in the spectral range 29.5--45~$\mu$m, with transitions up to 2939~K above the ground state. Many of the remaining weak features in the SWS spectrum can also be explained by including higher energy transitions of water. Here, we have analysed the LWS spectrum and carefully checked all of the prominent emission features. This shows that most of the detected lines can be attributed to rotational transitions in the vibrational ground state of water. 

\begin{table}
\begin{minipage}[t]{\columnwidth}
\caption{\label{nu2}Possible rotational lines from the vibrationally excited $\nu_{2}$=1 state of water in the LWS spectrum of VY CMa.}
\centering
\renewcommand{\footnoterule}{}  
\begin{tabular}{cccc}
\hline
\hline
Transition & Wavelength &  Energy of       & Flux\footnote{Errors do not include systematic uncertainty in the baseline subtraction ($\sim$30\%)}  \\
           & ($\mu$m)   &  Upper State (K) & ($10^{-19}$ W~cm$^{-2}$)\\
\hline
$3_{03}$--$2_{12}$  &  182.36 & 2491.8 & $1.3\pm0.3$ \\
$6_{33}$--$6_{24}$  &  140.06 & 3284.2 & $1.2\pm0.2$ \\
$4_{23}$--$4_{14}$  &  115.71 & 2745.2 & $2.3\pm0.8$ \\
\hline
\end{tabular}
\end{minipage}
\end{table}

The LWS spectra after continuum subtraction are shown in Fig.~\ref{vycmallines}. The LWS spectral resolution is lower than achieved with the SWS, and many of the features are blended together. However, at longer wavelengths there are at least some gaps with few or no water lines in at all, giving the possibility to uniquely identify several other molecular species. The continuum level was estimated individually for each detector by fitting a 3rd order polynomial through the lowest points in the spectrum as described in Sect.~\ref{obs}. For wavelengths greater than $\sim$80~$\mu$m, there are sufficient gaps between emission features, and this method should give a reasonable estimate of the continuum. However, severe line blending occurs $<60~\mu$m, and in this region the continuum level may have been overestimated because there is insufficiently clear space between the lines. This means that the line fluxes may be systematically underestimated. However, the fluxes for lines at wavelengths $<60~\mu$m are unlikely to be underestimated by more than a factor of a few, because sufficient contamination of the continuum would require very strong emission from lines lying $>$2000~K above the ground state and corresponding emission $>$80~$\mu$m is not seen.

In Fig.~\ref{vycmallines}, all the central wavelengths of all rotational ortho- and para- water transitions in its vibrational ground state are marked for energies up to 2000~K. Virtually every one of these transitions is matched by a feature in the data. This upper state energy is not a fundamental limit on the features detected in the spectrum, but adopted to ease of viewing the labels displayed at the LWS spectral resolution. The inclusion of lines with higher upper state energy levels suggests that many of the low lying lines are probably blended with other water features. Several spectral regions that are free of lines from transitions with energies below 2000~K clearly show that higher energy transitions are present. These wavelengths are individually marked in Fig.~\ref{vycmallines} (particularly 68--70~$\mu$m and 152--156~$\mu$m).

\subsection{Fitting the water lines \label{water_fit}}

In this section, we describe fitting the water lines in the vibrational ground state. Due to the large number of blended lines in the spectrum, any determination of the flux of individual lines is likely to be rather uncertain. In addition, there will be systematic errors introduced in the line fluxes due to the uncertainty in the underlying continuum level at the shortest wavelengths. In order to estimate the line fluxes, we have used a method that simultaneously fits the water lines in the vibrational ground state up to a certain energy level (upper state energy less than 2000~K above ground). The best fit is found by increasing the flux density at each line centre until the model reaches the data. The line wavelengths are fixed at their catalogue values \citep[taken from the JPL line catalogue;][]{pickett}, and line widths fixed at the spectral resolution of the ISO detectors \citep[0.29~$\mu$m for LWS detectors SW1--SW5 and 0.6~$\mu$m for detectors LW1--LW5;][]{gry}. Lines occuring within the wavelength range of each of the 10 LWS detectors are fitted together.

The flux densities at the centres of all lines measured by the detector are increased by a step that is weighted (for each line) relative to the Einstein coefficient for spontaneous emission, $A_{ul}$, \citep[with values taken from the JPL catalogue;][]{pickett}, and exp$(-E_{u}/kT_{weight})$, where $E_u/k$ is the energy of the upper state above ground in K, and $T_{weight}$ is a single temperature applied to all lines. This ensures that lines with higher weight are given preference in blends. However, it does not necessarily mean that the final flux will be distributed following the weights - the actual shape of the data may allow a line with a lower weight to have a higher flux than one with a high weight. The weighting scheme does not treat ortho and para lines separately. The basic step size in the model is set to 0.01 times flux density at the wavelength of the strongest line, and the individual step for all other lines are multiplied by their weights,
\begin{equation}
I_{\lambda-model}^{'} = I_{\lambda-model}+w_{\lambda}s
\end{equation}
where $I_{\lambda}$ is the flux density at the centre of the line, $w_{\lambda}$ is the weight for that line, defined by,
\begin{equation}\label{weight_full}
w_{\lambda}=\frac{A_{ul}}{{\rm{MAX}}[A_{ul}]}\frac{{\rm{exp}}(-E_{u}/kT_{weight})}{{\rm{MAX}}[{\rm{exp}}(-E_{u}/kT_{weight})]}
\end{equation}
and $s$ is the basic step size, defined by,
\begin{equation}
s = \frac{{\rm{MAX}}[I_{\lambda-data}]}{100}
\end{equation}
where the operation MAX finds the maximum value over the lines occuring within the wavelength range of the particular LWS detector being fitted.

A first estimate for the value of $T_{weight}$ to use in this weighting function was determined from a rotation diagram (see Sect.~\ref{excitation}) using a fit weighted relative to $A_{ul}$ only,
\begin{equation}\label{weight_Aul}
w_{\lambda}=\frac{A_{ul}}{{\rm{MAX}}[A_{ul}]}
\end{equation}

Figure~\ref{show_fit} shows a fairly representative example of the final fit, for the region 59--68~$\mu$m. This region clearly shows several higher energy lines that were not included in the fit.

The advantage of this technique to estimate the line fluxes is that it is much faster than carrying out a multiple Gaussian $\chi^{2}$ fit. Moreover, the flux in complicated blends is not overestimated (as it could be in a free Gaussian fit if the line shape is affected by other species). The disadvantage is that it takes no account of the overall excitation of the molecule, except within blends where the lines are explicitly weighted according to $A_{ul}$ and $T_{weight}$. The final errors on each line flux are highly correlated in the various blended groups and therefore difficult to rigorously estimate. However, this fitting method provides a good starting point to investigate the spectrum, and represents the best that can be achieved using data observed with the available spectral resolution.

In order to check the reliability of the automatic fitting routine, we compared the fit using several different weighting schemes: no weight, weighted relative to $A_{ul}$ only (equation~\ref{weight_Aul}), and weighted relative to $A_{ul}$ and exp$(-E_{u}/kT_{weight})$ (equation~\ref{weight_full}). For the few completely unblended lines, we also carried out a manual fit using ISAP and in general this agrees within the errors with the automatic procedure. The unblended line fluxes for these different fit methods are shown for VY CMa and IK Tau in Table~\ref{unblended}. We estimate that the fitting errors for {\it unblended} lines are 15--20\%. 

The final line fluxes determined from the fit for all lines up to 2000~K above the ground state (including blends) are shown in Tables~\ref{allfluxes1}, \ref{allfluxes2} and \ref{allfluxes3} for all of the stars investigated. In order to check the blended fits, we have compared our results for R Cas with those obtained by \citet{truongbach}. Our fitted values are in reasonable agreement (in a line by line comparison, many fluxes agree to better than 50\%, and the remaining discrepencies can be explained by the fact that we have included more blends). We estimate that our {\it blended} line fluxes are probably of the correct order of magnitude (unless the assumptions used for the weighting scheme within the fit are wildly incorrect). 

We have also checked our fitted fluxes with the analysis of \citet{maercker} who fitted water lines in a similar sample of stars. In general at long wavelengths where the lines are well separated, we find very good agreement in fitted fluxes. However, the difference in our analysis is that we have included the possibility of blends with higher energy lines. This is clearly necessary in VY CMa to account for all of the detected features. In the analysis of \citet{maercker}, they sometimes calculated model values lower than their quoted line fits to the data. This may be partly due to not identifying blends - some of our fitted fluxes which include blends (where Maercker et al. did not include a blend) give values closer to their model predictions. On the other hand, in some cases, our simple treatment that includes many blended lines seems to underestimate the flux compared to their model, suggesting that we may have not distributed the flux correctly. However, this comparison shows that in these stars, it is very important to consider that detected features may be due to blends of water transitions over a very wide range in energy, and it is not enough to merely consider the (strongest) lower level lines.

\begin{table*}
\begin{minipage}[t]{\linewidth}
\caption{\label{unblended}Fitted fluxes for the unblended lines in the spectrum of VY CMa and IK Tau.}
\centering
\renewcommand{\footnoterule}{}
\begin{tabular}{ccccc|c}
\hline \hline
Transition       & Wavelength & No Weight& Weighted  & Manual & IK Tau Weighted\\
                 &  ($\mu$m)  & Fit Flux &  Fit Flux&  Fit Flux\footnote{Errors do not include systematic uncertainty in the baseline subtraction ($\sim$30\%)}&  Fit Flux\\
                 &           & (10$^{-19}$ W cm$^{-2}$) & (10$^{-19}$ W cm$^{-2}$) & (10$^{-19}$ W cm$^{-2}$) & (10$^{-19}$ W cm$^{-2}$)\\
\hline
$4_{32}$--$3_{21}$ & 58.699  & 15.5 &  15.1  &   27.8$\pm$5.4  &     3.3   \\
$6_{16}$--$5_{05}$ & 82.031  & 10.9 &  11.0  &   14.0$\pm$2    &     2.5  \\
$2_{21}$--$1_{10}$ & 108.073 & 7.2  &  7.1   &   9.2$\pm$1.1   &     1.9  \\
$4_{32}$--$4_{23}$ & 121.722 & 3.1  &  3.0   &   3.8$\pm$1.0   &     0.73  \\
$4_{23}$--$4_{14}$ & 132.408 & 3.1  &  3.0   &   3.6$\pm$0.8   &     0.49  \\
$8_{36}$--$7_{43}$ & 133.549 & 2.2  &  2.1   &   2.7$\pm$0.8   &     0.54  \\
$5_{14}$--$5_{05}$ & 134.935 & 3.1  &  3.0   &   3.8$\pm$0.8   &     0.58  \\
$3_{30}$--$3_{21}$ & 136.496 & 2.9  &  2.8   &   3.3$\pm$0.9   &     0.73  \\
$4_{31}$--$4_{22}$ & 146.923 & 2.0  &  2.0   &   2.5$\pm$0.7   &     0.29  \\
$5_{32}$--$5_{23}$ & 160.510 & 2.0  &  1.9   &   2.5$\pm$0.4   &     0.55  \\
$2_{12}$--$1_{01}$ & 179.527 & 5.3  &  5.3   &   6.3$\pm$0.7   &     1.1  \\
$2_{21}$--$2_{12}$ & 180.488 & 2.6  &  2.6   &   3.4$\pm$0.4   &     0.85  \\
$6_{33}$--$5_{42}$ & 194.422 & 1.5  &  1.5   &   1.8$\pm$0.3   &     0.44  \\
\hline
\end{tabular}
\end{minipage}
\end{table*}

\subsection{Excitation of H$_2$O \label{excitation}}

In order to investigate the fitted line fluxes, we have used a simple rotation diagram approach \citep[e.g.][]{goldsmithRot}. Assuming that all transitions are in local thermodynamic equilibrium (LTE), there is a linear relationship between the natural logarithm of the upper state column density per statistical weight and the energy above the ground state (i.e. following the Boltzmann distribution),
\begin{equation}
{\rm{ln}}\left(\frac{N_u}{g_u}\right)={\rm{ln}}\left(\frac{N_{\rm{tot}}}{Q}\right)-\frac{1}{T}\frac{E_u}{k}
\end{equation}
where $N_{u}$ is the column density in the upper state, $g_u$ is the statistical weight of the upper state, $N_{\rm{tot}}$ is the total column density, and $Q$ is the partition function at temperature $T$. A plot of ln($N_u$/$g_u$) versus $E_u$/$k$ can be fitted with a straight line with gradient 1/$T$ and intercept ln($N_{\rm{tot}}$/$Q$). We have calculated the column densities assuming optically thin emission lines using the relationship,
\begin{equation}
N_{u}=\frac{4\pi\lambda}{hcA_{ul}\Omega}\int{I_{\lambda-data}\rm{d}\lambda}
\end{equation}
where $\Omega$ is the solid angle subtended by the water-emitting region around the star (assumed to be a disk with diameter 4$^{\prime\prime}$) and $\int{I_{\lambda-data}\rm{d}\lambda}$ is the measured flux. This choice of the source size is motivated by the size of the CO emission region imaged by \citet{muller}. Our adoption of a uniform temperature across a region of this size is highly unrealistic and so our results should be taken as qualitative estimates (also see further caveats on the results below).

\begin{figure}[!t]
\begin{center}
\resizebox{\hsize}{!}{\includegraphics{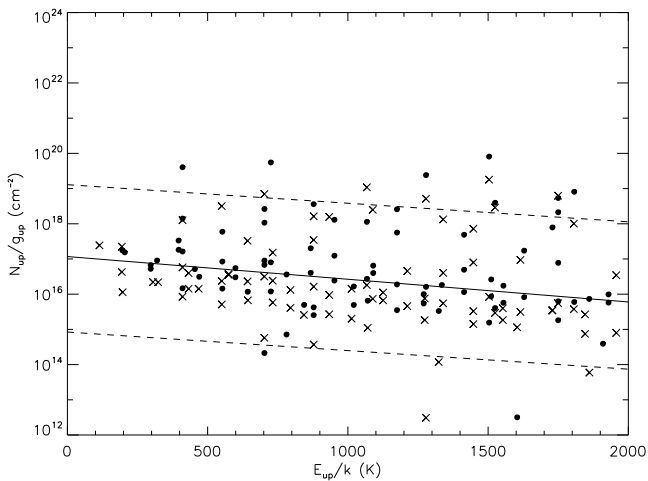}}
\end{center}
\caption{\label{water_rot}Rotation diagram resulting from fit to water lines in the LWS spectrum of VY CMa weighted by $A_{ul}$ and E$_{u}$. Ortho water lines are shown as filled circles and para water lines as crosses. The best fit to lines with upper state energies less than 2000 K (ignoring outlying points) is shown, giving a rotational temperature of 670$^{+210}_{-130}$~K.}
\end{figure}

Figure~\ref{water_rot} shows the rotation diagram for VY CMa. Even though we know that this model does not accurately describe reality (VY CMa actually has a very complex environment - see Sect.~\ref{summary}), and many of the lines are likely to be optically thick and/or subthermally excited, we think it provides an instructive starting point to view the fitted fluxes, particularly since the number of lines that we are fitting is very large (141 lines used in the fit). The lower energy lines in Fig.~\ref{water_rot} ($<1500$~K) are unlikely to be completely optically thin, and \citet{maercker} found that the water lines in their sample of stars were generally subthermally excited (although more thermalised in IK Tau than the other stars they looked at).

Although most of the results cluster along a straight line, there is a large scatter. This is exactly the effect we should expect to see for a nonlinear molecule which has a mixture of optically thick and thin lines that are not necessarily in LTE \citep[][see their Figs. 5 and 8]{goldsmithRot}. The lines least likely to be optically thick are those with upper state energies above 1500~K, although these lines are more likely to be blended, and so have large fitting errors.

Taking into account the above uncertainties, the fitted value for the total column density should be taken as a lower limit, and the temperature as a very rough estimate of the actual kinetic temperature (also bearing in mind that we expect a range of different thermal environments). The best fit straight line has a slope corresponding to a rotational temperature of $T_{\rm{rot}}$=670$^{+210}_{-130}$~K (the error was derived only taking into account the scatter in the data points). There does not appear to be a significant difference between the ortho- and para- lines. Due to the challenges described earlier regarding accurate flux estimation, we have discounted far outlying points by carrying out a 2$\sigma$ clip. These points, which can lie far above the best fit line, could be due to miss-fitting the flux of lines in blends. The rest of the scatter is due to real departures from the simple model as described above. More detailed modelling, beyond the scope of this paper is required to investigate these effects. The isotopic lines of H$_{2}^{18}$O could be used to investigate the optical depths, but this will require future higher resolution observations (a useful limit on the H$_{2}^{18}$O emission is difficult with the current LWS data).

The best fit intercept point gives $N_{\rm{tot}}/Q = (1.17\pm0.2)\times10^{17}$~cm$^{-2}$. The partition function at 300~K is given in the JPL catalogue \citep{pickett} as 178.12. The value scales as $T^{1.5}$ which leads to a partition function of 594.47 at 670~K, and this gives a lower limit on the total water column density of $(7.0\pm1.2)\times10^{19}$~cm$^{-2}$.

\subsection{Vibrationally excited water}

In addition to transitions from the vibrational ground state of water, \citet{neufeld_e} also identified 4 features with water in its vibrationally excited $\nu_{2}$=1 state. Rotational transitions of water in its $\nu_{2}$=1 state have also been observed at millimetre and submillimetre wavelengths, showing that the emission probably comes from a region close to the star \citep{menten_melnick,menten_vib}. At the spectral resolution of the LWS grating, vibrationally excited water is much harder to separate. However, in the longer wavelength detectors there are three features that match with wavelengths of $\nu_{2}$=1 rotational transitions. These are shown in Table~\ref{nu2}. It is possible that other vibrationally excited lines exist in the spectrum but all other transitions with similar energies occur in blends. Also, the ground state transitions were identified first, which may mean that some vibrationally excited lines could have been misidentified as ground state ones (for example, the $\nu_{2}$=1 $2_{12}-1_{01}$ line occurs at 170.928~$\mu$m, close to the feature we have labelled as $15_{511}-16_{214}$ in Fig.~\ref{vycmallines}). Higher spectral resolution observations would be required to unambiguously identify all of the vibrationally excited lines.

\citet{menten_vib} detected two transitions of vibrationally excited water using the 12m Atacama Pathfinder Experiment telescope (APEX): $5_{23}-6_{16}$ at 336.2~GHz and $6_{61}-7_{52}$ at 293.6~GHz. They used a model to calculate the optical depths and intensities of their measured lines, assuming that the transitions were thermalised. We have applied the same model to calculate the optical depth and predicted flux of our three detected FIR transitions. The line wavelengths and transition parameters were taken from the JPL line catalogue \citep{pickett}. If we make the same assumptions about the emitting region ($T$=1000~K, thermalised lines, uniform region within 0.05'' from the star) and assume a line width of 20~km~s$^{-1}$ (the same as the APEX lines), we derive high optical depths of 500, 270 and 600 for the three lines in Table~\ref{nu2}. The predicted fluxes are a few 10$^{-21}$ W~cm$^{-2}$, which underestimate our measured LWS fluxes by factors of 36--86. 

The model predictions were closer for the sub-millimetre lines observed by \citet{menten_vib}, underestimating their measured intensities by only a factor of 2.3 and 5, and a factor of 3 for the $5_{50}-6_{43}$ 232.6~GHz transition observed by \citet{menten_melnick}. The underestimated intensity was explained by the fact that the 294~GHz line may be boosted by weak maser action. 

We have also applied the model to the SWS lines observed by \citet{neufeld_e}, predicting that the SWS lines should be stronger than the LWS lines, which is not observed. The SWS fluxes measured by \citet{neufeld_e} are reproduced by the model within factors of 2--8. The optical depths for the SWS lines are predicted to be in the order of $10^4$. 

The model seems to be much worse at predicting the FIR fluxes than those in the sub-millimetre and mid-infrared. At the high optical depths predicted, the only parameters in the model that affect the final fluxes are the temperature of the emitting region, the line width and assumed source size. The flux calculated by the model could be brought into better agreement by a combination of increasing the temperature and size of the emitting region, as the line fluxes are $\sim$linearly proportional to both. However, increasing the assumed emission region size has a large effect on the predicted sub-mm main beam brightness temperatures. The excitation temperature does not have a large impact on the sub-mm lines, but a much higher temperature would be needed than is indicated by previous measurements \citep[see][]{menten_vib}.

This either indicates that there is another problem with the simple model (for example, different excitation conditions for the FIR lines) or that the measured LWS fluxes are overestimated (possibly due to additional line blending which has not been taken into account here). Observations at higher spectral resolution are needed to confirm the fluxes, and more detailed modelling is required, taking into account both the sub-mm and infrared lines.

\subsection{CO lines}

In the LWS spectrum of VY CMa, the water line density decreases toward the longer wavelengths. CO is known to be an abundant molecule in O-rich circumstellar envelopes and many observations have been made via its millimetre and sub-millimetre transitions \citep[e.g. ][]{kemper, decin, ziurys}. The lowest energy CO transition in the ISO spectral range is $J$=14--13, with upper state energy of 580~K above the ground state. 

Several CO lines have previously been reported in the LWS spectral range toward R Cas \citep{truongbach} and W Hya \citep{barlow}. In addition gaseous CO absorption has been observed around 4.5~$\mu$m in the SWS spectra of O-rich stars \citep[e.g.][]{sylvester}.

\begin{figure}[!t]
\begin{center}
\resizebox{\hsize}{!}{\includegraphics{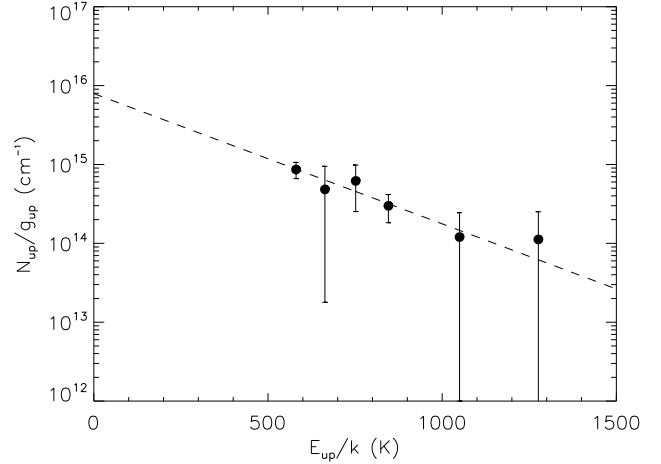}}
\end{center}
\caption{\label{co_rot}Rotation diagram for the CO lines observed towards VY CMa. The dashed line is a fit to the 4 lowest energy lines using the manually fitted fluxes from Table~\ref{co_lines}.}
\end{figure}

In the LWS spectrum of VY CMa, we observe some contribution to the spectrum from all CO lines from $J$=14--13 up to $J$=25--24. These lines are labeled in Fig.~\ref{vycmallines}. The higher energy lines are difficult to identify due to blending with water lines and the best detection is of the lowest energy lines which are generally well separated. The fluxes of these lines at long wavelengths were determined in two ways and are shown in Table~\ref{co_lines}. As a first estimate, the CO fluxes were determined by fitting the excess emission after subtracting the water line fit described in Sect.~\ref{water_fit}. The flux was also determined by a manual Gaussian fit using ISAP, taking account of nearby blended lines with a multi-Gaussian fit. The automatic fit generally underestimates the flux compared to the manual fit, probably because water was given preference in the blends for the automatic fit (although the automatic fit is within the errors of the ISAP fit). The true flux is probably between the two values.

We have carried out a rotation diagram analysis of the CO fluxes in the same way as described in Sect.~\ref{excitation} for H$_2$O. This is shown in Fig.~\ref{co_rot}, with a fit to the 4 lowest energy lines (the fit ignores the fluxes for $J=19-18$ and $J=21-20$ which have very large error bars). This gives a rough estimate of the rotational temperature and total column density as we are only fitting a few lines. However, the results should be more reliable than for H$_2$O, because the high excitation CO lines are likely to be optically thin. We have used the manual ISAP fluxes from Table~\ref{co_lines} and fitted a straight line weighted by their errors. This gives $T_{\rm{rot}}=260\pm140$~K and $N_{\rm{tot}}/Q = (8^{+22}_{-6})\times10^{15}$~cm$^{-2}$. The partition function at 300~K is given in the JPL catalogue \citep{pickett} as 108.865, and this leads to a total CO column density of $N_{\rm{tot}}=(9^{+24}_{-7})\times10^{17}$~cm$^{-2}$.

This column density is approximately 100 times lower than the H$_2$O column density calculated in Sect.~\ref{excitation}, showing that generally in the VY CMa envelope, water is the dominant oxygen carrier (although the H$_2$O and CO come from different temperature regions).

\begin{table}
\begin{minipage}[t]{\columnwidth}
\caption{\label{co_lines}Fitted fluxes for the CO lines observed above 120~$\mu$m for VY CMa.}
\centering
\renewcommand{\footnoterule}{}
\begin{tabular}{ccccc|c}
\hline \hline
Transition       & Wavelength & Auto & Manual  \\
                 &  ($\mu$m)  & Fit Flux &  Fit Flux\footnote{Errors do not include systematic uncertainty in the baseline subtraction ($\sim$30\%)}\\
                 &           & (10$^{-19}$ W cm$^{-2}$) & (10$^{-19}$ W cm$^{-2}$) \\
\hline
$J$=14--13 & 185.999 & 1.72  & 1.96$\pm$0.4  \\
$J$=15--14 & 173.631 & 1.35  & 1.88$\pm$1.3  \\
$J$=16--15 & 162.812 & 2.37  & 2.69$\pm$1.4  \\
$J$=17--16 & 153.267 & 1.54  & 2.00$\pm$0.6 \\
$J$=18--17 & 144.784 & -Blend- & -Blend-  \\
$J$=19--18 & 137.196 & 1.06  & 1.31$\pm$1.1  \\
$J$=20--19 & 130.369 & -Blend- & -Blend-  \\
$J$=21--20 & 124.193 & 1.60  & 0.92$\pm$2.0 \\
\hline
\end{tabular}
\end{minipage}
\end{table}

\subsection{Other lines: OH, NH$_{3}$ and H$_{3}$O$^{+}$}

Several other lines were detected in the spectrum of VY CMa. These are summarised in Table~\ref{other_lines} and described in the following paragraphs.

We detect strong emission from OH at 163~$\mu$m, which is blended with the CO $J$=16--15 line. This line is due to the lowest energy pure rotational transition in the $^{2}\Pi_{1/2}$ ladder of OH. Some contribution from the next transition in this ladder at 99~$\mu$m is probably also detected. In addition we detect emission in several other transitions of OH such as the lowest rotational line in the $^{2}\Pi_{3/2}$ ladder at 119~$\mu$m. As the cross ladder transition at 34~$\mu$m is seen in absorption in the SWS spectrum \citep{neufeld_e}, we might also expect the other cross ladder transitions in the LWS range to be in absorption (e.g. the other transitions from the $^{2}\Pi_{3/2}$ $J$=3/2 level at 53~$\mu$m and 79~$\mu$m). However, these are both blended with water emission.

The FIR lines of ammonia in its $\nu_{2}$ bending-inversion mode are rotation-inversion transitions, with the lowest energy transition $K$=2, $J$=3--2 at 165.60 and 169.97~$\mu$m. There is a clear emission peak at 165.60~$\mu$m in the spectrum of VY CMa (see Fig.~\ref{vycmallines}). This does not overlap with any water lines and has a width comparable to the instrument resolution. We think that the assignment of this feature to NH$_3$ is secure because it does not appear to be blended. There are only a few other species that could provide alternative explanations for the line: the most likely two choices are HNC $J$=20--19 and H$_{3}$O$^{+}$ $K$=0, $J$=5--4. The HNC line has a high energy upper level at 913~K and the corresponding transition in HCN at 161.35~$\mu$m is not clearly detected. If the line was due to H$_{3}$O$^{+}$, there should have been other features present which are not observed. The line is also observed in the spectrum of IK Tau (see Sect.~\ref{otherstars}).
 
The NH$_{3}$ line at 169.97~$\mu$m occurs in a blend with water lines. The transition at the next highest energy in NH$_{3}$ is $K$=3, $J$=4--3 at 124.65 and 127.11~$\mu$m. Both these lines occur in the wing of water emission (but are marked in Fig.~\ref{vycmallines}).

Ammonia has been observed in the envelopes of both carbon and oxygen rich stars - the first detection in an O-rich envelope was made by \citet{mclaren} in VY CMa, VX Sgr and IRC+10420. They observed several vibration-rotation transitions in the $\nu_{2}$ band at 10.5 and 10.7~$\mu$m. Subsequently these transitions have been measured in more detail for VY CMa by \citet[][]{monnier} and references therein. These observations show that NH$_{3}$ probably forms near the termination of the gas acceleration phase.

The only detection of the radio inversion transitions of ammonia in O-rich stars was made by \citet{menten_alcolea} toward IK Tau and IRC+10420. They found that the strongest inversion transition for IK Tau was $K$=3, $J$=3--3 whereas for IRC+10420 was $K$=1, $J$=1--1. This implies that with a higher mass loss rate, photodissociation occurs further from the star and the NH$_{3}$ emission comes from a lower temperature region.

The H$_{3}$O$^{+}$ ion has a similar pyramidal structure to the isoelectronic molecule NH$_{3}$. In an analogous way to NH$_{3}$, the oxygen atom can tunnel through the plane of the molecule leading to an inversion splitting of the rotational levels. However, for H$_{3}$O$^{+}$, the inversion splitting is very large \citep[55.34~cm$^{-1}$ for the $\nu_{2}$ bending-inversion mode;][]{liu_c}. This means that the pure inversion transitions occur at FIR wavelengths rather than in the radio regime as for ammonia. Several rotation-inversion transitions at 1~mm have previously been detected in giant molecular clouds \citep{wootten, phillips} and the lines near 300~GHz mapped toward Sgr~B2 \citep{vandertak}. Several of the FIR transitions have also been observed in absorption toward Sgr~B2 with ISO \citep{goicoechea_c,polehampton_survey}.

\begin{table}
\caption{\label{other_lines}Summary of other lines detected in the VY CMa spectrum.}
\begin{center}
\begin{tabular}{lcc}
\hline
\hline
Transition       & Wavelength & Comment \\
                 &  ($\mu$m)  & \\
\hline
OH $^{2}\Pi_{1/2}$ $J$=3/2--1/2 & 163.1/163.4 & blend with CO \\
OH $^{2}\Pi_{3/2}$ $J$=5/2--3/2 & 119.2/119.4 & well separated \\
NH$_3$ $K$=2, $J$=3$^{+}$--2$^{-}$ & 165.60 & well separated \\
NH$_3$ $K$=2, $J$=3$^{-}$--2$^{+}$ & 169.97 & blend with H$_2$O \\
H$_3$O$^+$ $K$=1, $J$=2--2 & 183.68 & possible identification \\
C$^+$ $^{2}$P$_{3/2}$--$^{2}$P$_{1/2}$ & 157.74 & blend with H$_2$O \\
U-line & 52.5 & \\ 
U-line & 86.6 & \\
\hline
\end{tabular}
\end{center}
\end{table}

The peak abundance of H$_{3}$O$^{+}$ in oxygen-rich circumstellar envelopes is predicted to be $\sim$10$^{-7}$ \citep{mamon}, with the major formation route being the photodissociation of H$_{2}$O and OH. Excitation is probably via the absorption of mid-IR radiation rather than collisions and thus absorption bands may also be observable at 10 and 17 $\mu$m. The only (possible) detection of H$_{3}$O$^{+}$ toward an evolved star so far is the 2$\sigma$ feature observed toward VY CMa by \citet{phillips} at the correct velocity for the $K$=0 $J$=3--2 396~GHz transition.

In the LWS grating spectrum of VY CMa, the spectral resolution is low and the inversion and rotation-inversion transitions are difficult to separate from the water lines. However, there is an unidentified feature at the correct wavelength for the $K$=1 $J$=2--2 transition at 183.68~$\mu$m. The other low energy inversion transition may contribute to the (water) emission features around 180--181~$\mu$m. The rotation-inversion transitions are harder to positively identify as they occur at shorter wavelengths where the density of water lines is higher.

In several O-rich star spectra, a feature at 157.8~$\mu$m has been assigned to the atomic fine structure line of ionised carbon at 157.74~$\mu$m \citep{truongbach,sylvester}. This line also appears in the VY CMa spectrum but no other atomic lines are clearly visible. \citet{sylvester} attribute this line to bad subtraction of the galactic background in their sources. However, at least some contribution to it can be explained by the 9$_{55}$--8$_{62}$ para water line at 157.88~$\mu$m. The upper level for this transition occurs 2031~K above the ground state. In VY CMa there could be some contribution from both lines. 

There are several remaining unassigned features in the spectrum and these are labeled as 'U-lines' in Fig.~\ref{vycmallines}. In addition, there are several species which have transitions in the range and are predicted to be reasonably abundant in O-rich circumstellar envelopes, but are not possible to detect due to line blending with water \citep[e.g. HCN, H$_{2}$S;][]{willacy}.

\begin{figure*}
\begin{center}
\resizebox{15.3cm}{!}{\includegraphics[angle=0]{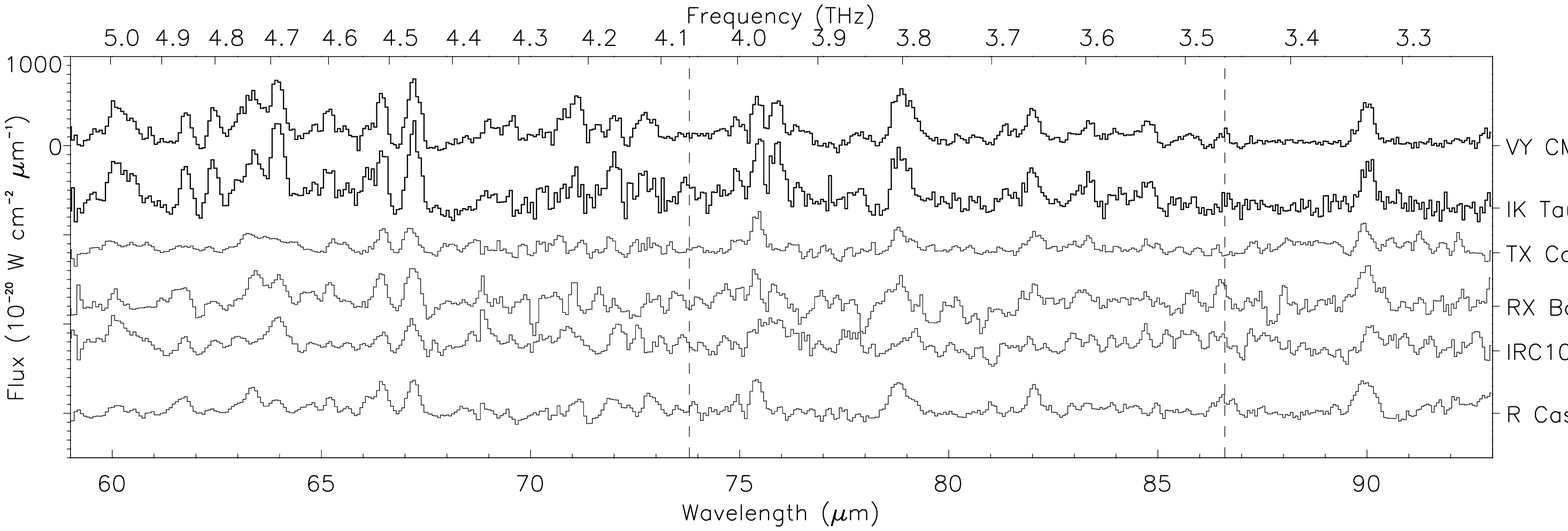}}
\resizebox{15.3cm}{!}{\includegraphics[angle=0]{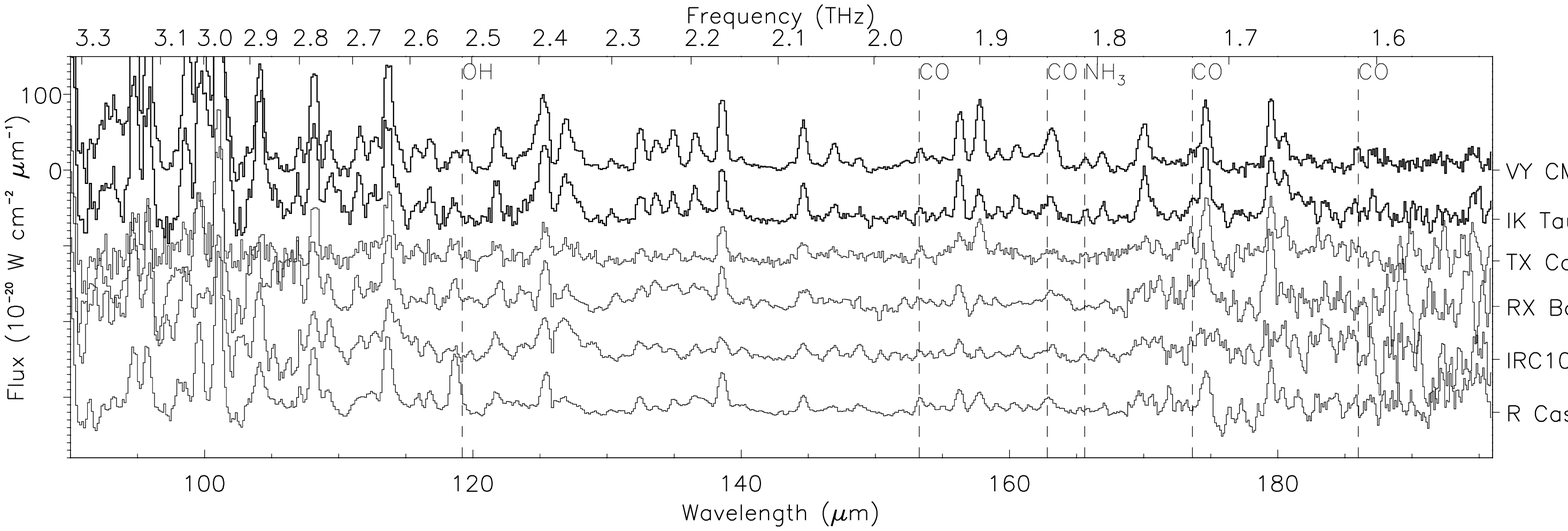}}
\end{center}
\caption{\label{compare}Comparison of the spectrum of VY CMa with the other stars (the spectra of the other stars have been smoothed and shifted for clarity - see text).}
\end{figure*}

\section{Comparison with other stars\label{otherstars}}

The spectrum of the oxygen rich Mira variable IK Tau shows remarkable similarity to that of VY CMa. However, the line fluxes shown in Fig.~\ref{compare} are $\sim$4 times weaker and the continuum level $\sim$10 times lower. Since it was only observed for about twice the integration time of VY CMa, the signal-to-noise of the spectrum appears lower than that of VY CMa. Within the noise level and accuracy of the continuum subtraction, all of the lines observed in the VY CMa spectrum are present in the IK Tau spectrum, generally with very similar relative intensities. Figure~\ref{compare} shows the spectrum of IK Tau scaled by a factor of 4.5 and compared to VY CMa. It is clear that the same lines are detected in each object. There are, however, a few exceptions where there are marked differences: there is much weaker OH emission with respect to water in IK Tau. The OH line at 119~$\mu$m is not present and there may be less contribution to the 163~$\mu$m CO/OH blend from OH. Also the line at 99~$\mu$m is weaker in IK Tau. However, the other lines such as CO, NH$_3$ and the possible $\nu_2=1$ water lines are all present in both stars. In addition, the unidentified feature at 86~$\mu$m is present in both spectra. There is only one feature that is present in IK Tau but not in VY CMa at 73.8~$\mu$m.

In Fig.~\ref{compare}, we also plot the other stars from Table~\ref{charcs}. In order to compare the spectral lines, we have
multiplied each continuum subtracted spectrum by factors of 4.5 for IK Tau and R Cas; 6 for TX Cam and 10 for RX Boo and IRC+10011. We have also smoothed the noisy spectra. Several key spectral lines are shown.

The other stars of a similar type to IK Tau (TX Cam, RX Boo, IRC+10011) have a much lower signal to noise ratio and their lines are weaker than IK Tau. However, the majority of detections in VY CMa and IK Tau also seem to be present in these stars.

We have fitted the water lines in the vibrational ground state for the other stars in the same way as for VY CMa and the results are shown in Tables~\ref{allfluxes1}, \ref{allfluxes2} and \ref{allfluxes3}. We have carried out a rotation diagram analysis for each star as described in Sect.~\ref{excitation}, and the results are shown in Table~\ref{all_rot}, giving a rough estimate of the rotational temperature and a lower limit for the total water column density.

The rotational temperatures for the other stars are very similar (within 25\%) to the value found for VY CMa, which indicates that the H$_2$O-emitting gas is excited under similar conditions. If the excitation is dominated by collisions, $T_{\rm{rot}}$ should reflect the kinetic temperature of the gas because of the high densities of AGB star envelopes. Alternatively, radiation may dominate the excitation, in which case $T_{\rm{rot}}$ would reflect the temperature of the ambient radiation field. The only star where $T_{\rm{rot}}$ deviates from the trend is IRC+10011, but in this case, the error on $T_{\rm{rot}}$ is rather large (due to a larger scatter in the data points in the rotation diagram).

The derived H$_2$O column densities for the other stars are lower than for VY CMa, with values ranging from 2\% to 12\% of the value for VY CMa. Since the other stars also have 10--100 times lower mass loss rates (see Table~\ref{charcs}), this result indicates that water is also the main oxygen carrier in the other star envelopes. However, a direct scaling of $N$(H$_2$O) with mass loss rate does not appear: the ratio of H$_2$O column density and mass loss rate shows a large variation over our sample of stars (factor of $>$10).

\begin{table}
\caption{\label{all_rot}Results of fitting a straight line to the rotation diagram for all of the stars in our sample.}
\begin{center}
\begin{tabular}{cccc}
\hline
\hline
Star      & $T_{\rm{rot}}$ & $N_{\rm{tot}}$(H$_2$O) & Number of Fitted \\
          &  (K)           & (10$^{18}$ cm$^{-2}$)  & Data Points \\
\hline
VY CMa    & 670$^{+210}_{-130}$   & $70\pm12$    & 141 \\
IK Tau    & 590$^{+150}_{-100}$   & $1.5\pm0.3$  & 140 \\
TX Cam    & 540$^{+140}_{-90}$    & $3.7\pm0.9$  & 125 \\
RX Boo    & 500$^{+110}_{-80}$    & $3.0\pm0.6$  & 125 \\
IRC+10011 & 1400$^{+1500}_{-500}$ & $5.6\pm1.3$  & 130 \\
R Cas     & 650$^{+270}_{-150}$   & $8.4\pm2.3$  & 132 \\
\hline
\end{tabular}
\end{center}
\end{table}

\section{Summary \label{summary}}

In this paper we present the ISO LWS spectrum of the luminous supergiant star VY CMa and give a brief description of the detected lines. We then compared the spectrum with several other evolved star spectra, particularly the Mira variable IK Tau.

The detected lines in the spectra of VY CMa and IK Tau are remarkably similar, both in the species detected and the relative intensity of the water lines.

We draw the following conclusions:

\begin{itemize}
\item We confirm that the spectra of our sample of evolved stars are dominated by water line emission in the FIR, and we report the detection of nearly all the H$_2$O lines up to $\sim$ 2000~K above the ground state.
\item A simple rotation analysis of the water lines shows that there are a probably a mixture of optically thick and thin lines which may be subthermally excited (the fit gives a rotational temperature of 670$^{+210}_{-130}$~K, and a corresponding column density $N_{\rm{tot}}=(7.0\pm1.2)\times10^{19}$~cm$^{-2}$ which should be treated a lower limit to the true value).
\item We tentatively assign several of the detected features to excited water in its $\nu_2=1$ vibrational state.
\item We estimate the column density of CO to be $(9^{+24}_{-7})\times10^{17}$~cm$^{-2}$, showing that water is the dominant oxygen carrier in these envelopes.
\item We present a detection of ammonia in VY CMa and IK Tau, and a tentative detection of the H$_3$O$^+$ ion toward VY CMa.
\end{itemize}

We note that the kinematics around VY CMa are very complex as shown by interferometric CO observations \citep{muller} which suggest the presence of spherically symmetric outflow together with a high velocity bipolar flow. The water maser observations by \citet{menten} further imply that some fraction of the water emission may be emitted by special regions associated with shock fronts, excited in photochemical reactions associated with the outflow material. The presence of this complex excitation environment has also been indicated by observations of other molecules \citep{ziurys, smith}. To bring together an accurate understanding of the local environment will require a sophisticated model beyond the scope of the present study. Future observations with Herschel and ALMA should bring the necessary data to understand this complicated source. In particular, HIFI will add the ground state water lines, the isotopic H$_2^{18}$O lines and the line profiles which may be used to study kinematics and self-absorption effects, and PACS will reobserve all of the lines above 55~$\mu$m. ALMA will give the spatial distribution of H$_2$O and other molecules in these AGB envelopes.

\begin{acknowledgements}
E. T. P. would like to thank E. Kr\"{u}gel for help and encouragement in Bonn, and also M. Barlow for helpful advice and T. W. Grundy for useful discussions about the ISO L01 spectra and HPDP data. We made use of the Cologne Database for Molecular Spectroscopy \citep[CDMS; ][]{mueller}. The LWS Interactive Analysis (LIA) package is a joint development of the ISO-LWS Instrument Team at the Rutherford Appleton Laboratory (RAL, UK - the PI Institute) and the Infrared Processing and Analysis Center (IPAC/Caltech, USA). The ISO Spectral Analysis Package (ISAP) is a joint development by the LWS and SWS Instrument Teams and Data Centres. Contributing institutes are CESR, IAS, IPAC, MPE, RAL and SRON.
\end{acknowledgements}

\bibliographystyle{aa}
\bibliography{13310.bbl}

\appendix
\section{Full list of fitted fluxes}

\begin{table*}
\caption{\label{allfluxes1}All line fluxes calculated by our fitting procedure for the 6 stars.}
\begin{center}
\begin{tabular}{ccccccccc}
\hline
\hline
Transition           & Wavelength & Energy of& VY CMa & IK Tau & TX Cam & RX Boo & IRC+10011 & R Cas\\
                     & ($\mu$m) & Upper& ($10^{-16}$ W m$^{-2}$) & & & & &\\
                     &          & State (K) &  &  &  &  &  & \\
\hline
SW1   &    &    &    &    &    &    &    &   \\
$6_{60}$--$7_{35}$       &  43.790  &  1175.0  &  146  &  29.7  &  7.2  &  5.0  &  18.1  &  11.1 \\
$5_{41}$--$4_{32}$       &  43.893  &  550.4   &  49.2  &  11.5  &  1.3  &  5.6  &  3.6  &  5.4 \\
$7_{43}$--$7_{16}$       &  44.048  &  1013.2  &  208  &  49.9  &  7.9  &  21.1  &  21.7  &  17.0 \\
$5_{42}$--$4_{31}$       &  44.195  &  552.3   &  183  &  48.7  &  17.7  &  6.2  &  12.1  &  25.2 \\
$8_{36}$--$7_{25}$       &  44.702  &  1125.7  &  174  &  46.1  &  7.3  &  9.7  &  19.1  &  7.6 \\
$5_{23}$--$4_{14}$       &  45.112  &  323.5   &  323  &  102  &  18.8  &  33.4  &  19.0  &  32.5 \\
$3_{31}$--$2_{02}$       &  46.484  &  100.8   &  200  &  52.3  &  11.4  &  13.8  &  19.7  &  27.4 \\
$6_{42}$--$6_{15}$       &  46.539  &  781.1   &  55.3  &  22.4  &  2.2  &  5.7  &  5.0  &  5.2 \\
$6_{33}$--$6_{06}$       &  46.544  &  642.7   &  13.0  &  4.4  &  0.49  &  1.1  &  0.82  &  1.1 \\
$10_{29}$--$9_{18}$      &  46.608  &  1552.6  &  13.2  &  5.5  &  3.4  &  0.83  &  0  &  5.6 \\
$7_{35}$--$6_{24}$       &  46.746  &  867.3   &  102  &  17.4  &  10.1  &  0.11  &  2.4  &  17.7 \\
$4_{40}$--$4_{13}$       &  47.028  &  396.4   &  63.1  &  14.1  &  4.5  &  6.3  &  2.0  &  4.8 \\
$10_{19}$--$9_{28}$      &  47.029  &  1554.4  &  4.0  &  1.0  &  0.26  &  0.47  &  0.24  &  0.30 \\
$11_{111}$--$10_{010}$   &  47.040  &  1603.6  &  0  &  0  &  0  &  0  &  0  &  0 \\
$11_{011}$--$10_{110}$   &  47.045  &  1603.6  &  117  &  30.0  &  7.5  &  14.9  &  8.2  &  9.4 \\
$8_{44}$--$9_{19}$       &  47.267  &  1324.0  &  0  &  0  &  1.3  &  0  &  0  &  0 \\
$5_{41}$--$5_{14}$       &  47.420  &  574.7   &  312  &  62.4  &  24.5  &  27.0  &  21.2  &  43.8 \\
$7_{61}$--$8_{36}$       &  47.601  &  1447.6  &  115  &  16.1  &  8.1  &  16.4  &  2.0  &  17.2 \\
$5_{32}$--$4_{23}$       &  47.973  &  432.2   &  304  &  64.4  &  23.7  &  23.1  &  11.2  &  37.7 \\
$4_{40}$--$3_{31}$       &  49.282  &  410.4   &  6.2  &  0.12  &  0.01  &  4.1  &  0.02  &  0.02 \\
$6_{61}$--$7_{34}$       &  49.334  &  1212.0  &  77.2  &  25.8  &  2.2  &  6.8  &  5.9  &  7.2 \\
$4_{41}$--$3_{30}$       &  49.337  &  410.7   &  50.3  &  15.2  &  1.6  &  3.9  &  2.3  &  4.8 \\
$6_{34}$--$5_{23}$       &  49.391  &  642.4   &  167  &  50.7  &  13.3  &  9.4  &  4.5  &  15.7 \\
$9_{28}$--$8_{17}$       &  50.634  &  1270.3  &  260  &  77.3  &  9.9  &  10.1  &  16.4  &  13.8 \\
\hline
SW2   &    &    &    &    &    &    &    &   \\
$9_{27}$--$8_{36}$       &  51.071  &  1447.6  &  267  &  62.1  &  15.1  &  17.5  &  9.2  &  12.8 \\
$10_{110}$--$9_{09}$     &  51.445  &  1323.9  &  217  &  51.2  &  0.62  &  10.2  &  10.9  &  4.8 \\
$10_{010}$--$9_{19}$     &  51.461  &  1324.0  &  0.20  &  0.05  &  0  &  1.0  &  0.01  &  1.7 \\
$9_{18}$--$8_{27}$       &  51.685  &  1274.2  &  230  &  41.0  &  18.5  &  8.4  &  6.3  &  3.1 \\
$5_{50}$--$6_{25}$       &  52.864  &  795.5  &  44.1  &  14.1  &  13.7  &  7.9  &  0  &  17.2 \\
$5_{33}$--$4_{22}$       &  53.138  &  454.3  &  175  &  37.7  &  23.3  &  12.1  &  3.4  &  11.1 \\
$7_{43}$--$8_{18}$       &  53.455  &  1070.7  &  24.9  &  12.8  &  29.3  &  10.9  &  0.70  &  8.3 \\
$5_{32}$--$5_{05}$       &  54.507  &  468.1  &  158  &  60.6  &  21.6  &  9.2  &  16.0  &  14.5 \\
$8_{27}$--$7_{16}$       &  55.131  &  1013.2  &  149  &  34.2  &  0.62  &  26.6  &  12.6  &  9.8 \\
$10_{29}$--$10_{110}$    &  55.840  &  1603.6  &  0  &  0.40  &  0  &  6.6  &  2.0  &  0 \\
$6_{51}$--$7_{26}$       &  55.858  &  1021.0  &  104  &  42.0  &  0.54  &  12.9  &  3.8  &  0 \\
$10_{19}$--$10_{010}$    &  56.027  &  1603.6  &  66.8  &  1.9  &  1.5  &  0.03  &  0  &  0 \\
$4_{31}$--$3_{22}$       &  56.325  &  296.8  &  155  &  27.4  &  0  &  14.9  &  0  &  0 \\
$9_{19}$--$8_{08}$       &  56.771  &  1070.5  &  126  &  12.9  &  0.78  &  13.6  &  0  &  0 \\
$9_{09}$--$8_{18}$       &  56.816  &  1070.7  &  13.4  &  3.9  &  0.90  &  0.94  &  0  &  0.78 \\
$8_{53}$--$9_{28}$       &  56.972  &  1554.4  &  109  &  20.4  &  0.15  &  11.1  &  4.6  &  12.4 \\
$7_{52}$--$8_{27}$       &  57.394  &  1274.2  &  73.4  &  9.2  &  7.2  &  6.5  &  0  &  9.7 \\
$4_{22}$--$3_{13}$       &  57.637  &  204.7  &  142  &  29.4  &  10.6  &  21.6  &  6.1  &  14.6 \\
$8_{17}$--$7_{26}$       &  57.709  &  1021.0  &  221  &  45.8  &  10.4  &  16.4  &  13.9  &  19.0 \\
$6_{42}$--$7_{17}$       &  58.377  &  843.8  &  0  &  0.20  &  0  &  5.3  &  0  &  0 \\
$4_{32}$--$3_{21}$       &  58.699  &  305.2  &  151  &  33.4  &  7.6  &  1.5  &  11.7  &  15.0 \\
\hline
SW3   &    &    &    &    &    &    &    &   \\
$7_{26}$--$6_{15}$   &  59.987  &  781.1  &  76.4  &  17.7  &  3.8  &  4.8  &  8.0  &  1.8 \\
$8_{26}$--$7_{35}$   &  60.162  &  1175.0  &  106  &  28.4  &  4.8  &  2.5  &  7.5  &  7.2 \\
$7_{62}$--$8_{35}$   &  60.229  &  1511.0  &  8.4  &  0.90  &  0.08  &  0  &  1.8  &  0.18 \\
$5_{41}$--$6_{16}$   &  61.316  &  643.5  &  10.5  &  0  &  0  &  0  &  0  &  1.0 \\
$4_{31}$--$4_{04}$   &  61.809  &  319.5  &  95.2  &  24.8  &  4.9  &  4.5  &  2.5  &  11.7 \\
$4_{40}$--$5_{15}$   &  61.916  &  469.9  &  1.9  &  0.49  &  0  &  0.06  &  0.05  &  0.06 \\
$9_{36}$--$8_{45}$   &  62.418  &  1615.3  &  10.7  &  3.6  &  0.46  &  0.31  &  0.28  &  0.49 \\
$9_{28}$--$9_{19}$   &  62.432  &  1324.0  &  103  &  27.8  &  2.0  &  0.41  &  2.9  &  1.4 \\
$9_{18}$--$9_{09}$   &  62.928  &  1324.0  &  68.6  &  16.4  &  1.8  &  0  &  5.5  &  6.2 \\
$8_{18}$--$7_{07}$   &  63.324  &  843.5  &  71.5  &  13.1  &  3.9  &  2.9  &  0  &  11.7 \\
$8_{08}$--$7_{17}$   &  63.458  &  843.8  &  139  &  35.7  &  7.2  &  11.6  &  3.6  &  13.7 \\
$7_{62}$--$7_{53}$   &  63.880  &  1524.6  &  12.1  &  2.7  &  0.01  &  0.01  &  0.23  &  0.19 \\
$6_{61}$--$6_{52}$   &  63.914  &  1278.5  &  88.9  &  25.3  &  3.1  &  6.2  &  4.5  &  5.0 \\
$6_{60}$--$6_{51}$   &  63.928  &  1278.6  &  5.5  &  1.6  &  0.20  &  0.34  &  0.30  &  0.30 \\
$7_{61}$--$7_{52}$   &  63.955  &  1524.9  &  110  &  32.4  &  4.3  &  5.2  &  6.0  &  5.0 \\
$6_{25}$--$5_{14}$   &  65.166  &  574.7  &  107  &  24.2  &  1.3  &  8.3  &  3.6  &  10.1 \\
$7_{16}$--$6_{25}$   &  66.093  &  795.5  &  60.3  &  27.2  &  2.7  &  2.2  &  0.84  &  8.6 \\
$3_{30}$--$2_{21}$   &  66.438  &  194.1  &  156  &  37.8  &  10.3  &  11.8  &  5.2  &  20.7 \\
$3_{31}$--$2_{20}$   &  67.089  &  195.9  &  87.7  &  20.9  &  11.0  &  7.0  &  7.7  &  8.1 \\
$3_{30}$--$3_{03}$   &  67.269  &  196.8  &  153  &  41.7  &  6.3  &  10.3  &  5.5  &  18.5 \\
\hline
\end{tabular}
\end{center}
\end{table*}

\begin{table*}
\caption{\label{allfluxes2}Line fluxes for all stars, continued.}
\begin{center}
\begin{tabular}{ccccccccc}
\hline
\hline
Transition           & Wavelength & Energy of& VY CMa & IK Tau & TX Cam & RX Boo & IRC+10011 & R Cas\\
                     & ($\mu$m) & Upper& ($10^{-16}$ W m$^{-2}$) & & & & &\\
                     &          & State (K) &  &  &  &  &  & \\
\hline
SW4   &    &    &    &    &    &    &    &   \\
$8_{27}$--$8_{18}$   &  70.703  &  1070.7  &  81.7  &  7.5  &  6.6  &  3.6  &  7.7  &  2.2 \\
$5_{24}$--$4_{13}$   &  71.067  &  396.4  &  145  &  25.9  &  3.2  &  8.3  &  3.0  &  3.6 \\
$7_{17}$--$6_{06}$   &  71.540  &  642.7  &  56.2  &  5.8  &  1.2  &  3.4  &  0  &  0 \\
$5_{51}$--$6_{24}$   &  71.788  &  867.3  &  1.10  &  0.57  &  0  &  0  &  0.09  &  2.1 \\
$7_{07}$--$6_{16}$   &  71.947  &  643.5  &  85.4  &  30.4  &  5.4  &  2.2  &  4.4  &  10.6 \\
$8_{17}$--$8_{08}$   &  72.032  &  1070.5  &  18.3  &  10.3  &  0  &  0.81  &  6.2  &  0 \\
$9_{37}$--$9_{28}$   &  73.613  &  1554.4  &  37.1  &  15.8  &  0  &  2.0  &  6.6  &  1.4 \\
$7_{25}$--$6_{34}$   &  74.945  &  933.7  &  81.0  &  25.6  &  3.7  &  5.3  &  6.1  &  12.7 \\
$3_{21}$--$2_{12}$   &  75.381  &  114.4  &  94.8  &  23.0  &  10.9  &  10.2  &  2.3  &  14.7 \\
$8_{54}$--$8_{45}$   &  75.496  &  1615.3  &  68.2  &  30.6  &  6.6  &  3.8  &  6.5  &  12.7 \\
$5_{51}$--$5_{42}$   &  75.781  &  877.8  &  47.2  &  13.0  &  0.69  &  6.7  &  5.0  &  0 \\
$7_{53}$--$7_{44}$   &  75.813  &  1334.8  &  18.7  &  4.7  &  0.16  &  2.9  &  1.7  &  0 \\
$6_{52}$--$6_{43}$   &  75.830  &  1088.8  &  0.03  &  0.01  &  0  &  0  &  0  &  0 \\
$5_{50}$--$5_{41}$   &  75.910  &  878.2  &  93.5  &  33.8  &  2.5  &  0.84  &  5.3  &  5.2 \\
$6_{51}$--$6_{42}$   &  76.422  &  1090.3  &  50.9  &  15.9  &  0  &  0  &  3.9  &  5.3 \\
$7_{52}$--$7_{43}$   &  77.761  &  1339.9  &  37.4  &  10.1  &  0  &  4.2  &  3.9  &  0 \\
\hline
SW5   &    &    &    &    &    &    &    &   \\
$4_{23}$--$3_{12}$   &  78.742  &  249.4  &  110  &  23.2  &  11.2  &  4.0  &  1.7  &  12.6 \\
$6_{15}$--$5_{24}$   &  78.928  &  598.8  &  125  &  26.1  &  4.8  &  7.8  &  4.7  &  9.8 \\
$9_{46}$--$9_{37}$   &  80.222  &  1749.9  &  32.3  &  6.3  &  2.9  &  5.2  &  0  &  0 \\
$8_{53}$--$8_{44}$   &  80.557  &  1628.4  &  28.7  &  6.2  &  1.1  &  0  &  0  &  0 \\
$7_{26}$--$7_{17}$   &  81.216  &  843.8  &  34.0  &  4.8  &  2.0  &  0  &  0.84  &  0.32 \\
$9_{27}$--$9_{18}$   &  81.405  &  1552.6  &  43.2  &  10.2  &  1.7  &  0  &  0.06  &  0 \\
$8_{35}$--$7_{44}$   &  81.690  &  1334.8  &  27.6  &  2.4  &  0.09  &  4.1  &  2.0  &  0 \\
$6_{16}$--$5_{05}$   &  82.031  &  468.1  &  110  &  24.8  &  8.8  &  9.4  &  1.4  &  16.0 \\
$8_{36}$--$8_{27}$   &  82.977  &  1274.2  &  35.3  &  9.0  &  5.0  &  4.6  &  5.6  &  6.8 \\
$6_{06}$--$5_{15}$   &  83.284  &  469.9  &  60.0  &  17.0  &  4.9  &  3.7  &  0.50  &  6.0 \\
$7_{16}$--$7_{07}$   &  84.767  &  843.5  &  75.2  &  16.9  &  5.1  &  6.7  &  7.6  &  3.8 \\
$8_{45}$--$8_{36}$   &  85.769  &  1447.6  &  36.5  &  3.4  &  0  &  4.1  &  5.1  &  0 \\
$3_{22}$--$2_{11}$   &  89.988  &  136.9  &  85.4  &  19.0  &  8.1  &  7.5  &  2.9  &  10.0 \\
$7_{44}$--$7_{35}$   &  90.050  &  1175.0  &  50.1  &  10.8  &  2.0  &  10.4  &  5.3  &  11.8 \\
\hline
LW1   &    &    &    &    &    &    &    &   \\
$6_{43}$--$6_{34}$     &  92.811  &  933.7  &  43.2  &  5.2  &  0  &  3.0  &  3.5  &  0 \\
$7_{35}$--$7_{26}$     &  93.383  &  1021.0  &  40.3  &  4.6  &  1.1  &  0  &  4.2  &  3.0 \\
$6_{52}$--$7_{25}$     &  94.172  &  1125.7  &  32.8  &  5.0  &  1.7  &  0  &  3.9  &  1.8 \\
$5_{42}$--$5_{33}$     &  94.210  &  725.1  &  3.3  &  0.66  &  1.2  &  0  &  0.35  &  0.20 \\
$6_{25}$--$6_{16}$     &  94.644  &  643.5  &  43.4  &  8.1  &  1.3  &  1.9  &  2.3  &  2.9 \\
$4_{41}$--$4_{32}$     &  94.705  &  550.4  &  65.3  &  13.7  &  3.0  &  3.7  &  4.0  &  7.3 \\
$9_{45}$--$8_{54}$     &  95.176  &  1805.9  &  46.6  &  9.3  &  1.9  &  3.7  &  2.9  &  1.8 \\
$5_{15}$--$4_{04}$     &  95.627  &  319.5  &  73.7  &  14.0  &  3.4  &  4.9  &  4.5  &  10.8 \\
$4_{40}$--$4_{31}$     &  95.885  &  552.3  &  59.1  &  13.6  &  3.9  &  2.3  &  3.1  &  2.8 \\
$9_{37}$--$10_{010}$   &  98.329  &  1603.6  &  45.8  &  9.3  &  2.1  &  2.3  &  3.7  &  2.3 \\
$5_{41}$--$5_{32}$     &  98.494  &  732.1  &  54.8  &  7.7  &  1.1  &  0.56  &  3.7  &  3.2 \\
$5_{05}$--$4_{14}$     &  99.493  &  323.5  &  85.2  &  18.8  &  6.4  &  6.0  &  3.9  &  15.4 \\
$8_{26}$--$8_{17}$     &  99.979  &  1270.3  &  92.4  &  22.0  &  2.8  &  3.6  &  3.9  &  6.1 \\
$5_{14}$--$4_{23}$     &  100.913  &  432.2  &  89.3  &  23.6  &  6.0  &  5.5  &  3.7  &  11.2 \\
$2_{20}$--$1_{11}$     &  100.983  &  53.4  &  105  &  25.4  &  5.9  &  7.2  &  3.9  &  14.3 \\
$6_{24}$--$5_{33}$     &  101.209  &  725.1  &  84.4  &  21.2  &  5.4  &  3.5  &  2.8  &  8.7 \\
$6_{42}$--$6_{33}$     &  103.916  &  951.8  &  52.8  &  15.7  &  2.4  &  4.3  &  3.6  &  5.9 \\
$6_{15}$--$6_{06}$     &  103.940  &  642.7  &  0.56  &  0.17  &  0.03  &  0.85  &  0.03  &  0.07 \\
$6_{34}$--$6_{25}$     &  104.094  &  795.5  &  35.5  &  8.1  &  3.4  &  0.85  &  1.7  &  3.7 \\
\hline
\end{tabular}
\end{center}
\end{table*}

\begin{table*}
\caption{\label{allfluxes3}Line fluxes for all stars, continued.}
\begin{center}
\begin{tabular}{ccccccccc}
\hline
\hline
Transition           & Wavelength & Energy of& VY CMa & IK Tau & TX Cam & RX Boo & IRC+10011 & R Cas\\
                     & ($\mu$m) & Upper& ($10^{-16}$ W m$^{-2}$) & & & & &\\
                     &          & State (K) &  &  &  &  &  & \\
\hline
LW2   &    &    &    &    &    &    &    &   \\
$2_{21}$--$1_{10}$     &  108.073  &  61.0  &  70.6  &  18.2  &  9.7  &  6.3  &  4.1  &  15.3 \\
$5_{24}$--$5_{15}$     &  111.628  &  469.9  &  30.0  &  8.5  &  3.6  &  0.74  &  2.2  &  3.4 \\
$7_{43}$--$7_{34}$     &  112.511  &  1212.0  &  21.8  &  5.1  &  1.5  &  1.9  &  2.1  &  4.0 \\
$4_{41}$--$5_{14}$     &  112.803  &  574.7  &  12.7  &  3.0  &  0.89  &  1.7  &  1.6  &  1.7 \\
$4_{14}$--$3_{03}$     &  113.537  &  196.8  &  59.2  &  15.7  &  5.1  &  3.6  &  3.0  &  13.3 \\
$5_{33}$--$5_{24}$     &  113.948  &  598.8  &  59.1  &  13.8  &  5.5  &  5.1  &  4.1  &  10.7 \\
$9_{27}$--$10_{110}$   &  114.454  &  1603.6  &  11.7  &  5.1  &  0.76  &  1.1  &  2.8  &  1.9 \\
$8_{36}$--$9_{09}$     &  116.350  &  1323.9  &  10.2  &  2.2  &  0  &  0.04  &  0.92  &  0.19 \\
$7_{34}$--$6_{43}$     &  116.779  &  1088.8  &  21.3  &  6.0  &  0.35  &  2.0  &  1.4  &  4.8 \\
$9_{46}$--$8_{53}$     &  117.684  &  1807.0  &  1.7  &  0  &  0  &  0.19  &  0.52  &  0.36 \\
$9_{37}$--$8_{44}$     &  118.405  &  1628.4  &  11.0  &  3.1  &  1.2  &  1.3  &  0.81  &  10.7 \\
$4_{32}$--$4_{23}$     &  121.722  &  432.2  &  30.0  &  7.3  &  1.7  &  1.7  &  2.8  &  4.5 \\
$8_{44}$--$8_{35}$     &  122.522  &  1511.0  &  10.2  &  2.2  &  0.27  &  0  &  0.55  &  1.8 \\
$9_{36}$--$9_{27}$     &  123.460  &  1729.3  &  10.7  &  2.0  &  0.22  &  1.5  &  1.7  &  1.8 \\
$4_{04}$--$3_{13}$     &  125.354  &  204.7  &  51.8  &  15.3  &  4.3  &  4.0  &  4.0  &  7.3 \\
\hline
LW3   &    &    &    &    &    &    &    &   \\
$3_{31}$--$3_{22}$   &  126.714  &  296.8  &  33.0  &  8.9  &  3.5  &  0.76  &  4.1  &  4.4 \\
$7_{25}$--$7_{16}$   &  127.884  &  1013.2  &  14.8  &  4.0  &  2.7  &  0.78  &  2.4  &  1.5 \\
$9_{45}$--$9_{36}$   &  129.339  &  1845.8  &  3.0  &  0.38  &  0.61  &  0  &  1.4  &  0 \\
$7_{53}$--$8_{26}$   &  130.319  &  1414.2  &  8.4  &  2.3  &  1.0  &  0.85  &  0.45  &  0.41 \\
$4_{23}$--$4_{14}$   &  132.408  &  323.5  &  30.1  &  4.9  &  2.5  &  2.4  &  1.8  &  4.3 \\
$8_{36}$--$7_{43}$   &  133.549  &  1339.9  &  21.0  &  5.4  &  2.0  &  3.3  &  1.6  &  1.7 \\
$5_{14}$--$5_{05}$   &  134.935  &  468.1  &  30.3  &  5.8  &  2.0  &  2.7  &  0.99  &  3.5 \\
$3_{30}$--$3_{21}$   &  136.496  &  305.2  &  27.6  &  7.3  &  2.8  &  3.1  &  1.6  &  4.0 \\
$7_{35}$--$8_{08}$   &  137.683  &  1070.5  &  6.6  &  2.1  &  0.92  &  1.4  &  1.2  &  1.8 \\
$3_{13}$--$2_{02}$   &  138.528  &  100.8  &  45.1  &  9.3  &  6.0  &  5.4  &  1.8  &  8.1 \\
$8_{44}$--$7_{53}$   &  138.641  &  1524.6  &  11.4  &  2.4  &  1.1  &  1.0  &  0.53  &  2.1 \\
$4_{13}$--$3_{22}$   &  144.518  &  296.8  &  32.2  &  7.4  &  3.1  &  2.4  &  1.4  &  3.7 \\
$4_{31}$--$4_{22}$   &  146.923  &  454.3  &  19.7  &  2.9  &  2.3  &  0.9  &  1.3  &  1.5 \\
\hline
LW4   &    &    &    &    &    &    &    &   \\
$8_{35}$--$8_{26}$   &  148.708  &  1414.2  &  5.6  &  0.83  &  0.65  &  1.4  &  0.73  &  0.71 \\
$5_{42}$--$6_{15}$   &  148.790  &  781.1  &  3.6  &  1.4  &  0.29  &  0.22  &  0.65  &  0.69 \\
$3_{22}$--$3_{13}$   &  156.194  &  204.7  &  5.7  &  1.5  &  0.75  &  0.67  &  0.21  &  0.54 \\
$5_{23}$--$4_{32}$   &  156.265  &  550.4  &  37.0  &  10.0  &  2.9  &  1.7  &  1.6  &  3.4 \\
$3_{31}$--$4_{04}$   &  158.312  &  319.5  &  13.4  &  0.48  &  1.6  &  0.33  &  0.19  &  0.83 \\
$8_{45}$--$7_{52}$   &  159.051  &  1524.9  &  10.4  &  1.8  &  0.81  &  0.59  &  0.73  &  1.2 \\
$6_{34}$--$7_{07}$   &  159.400  &  843.5  &  6.8  &  1.2  &  0.79  &  0.20  &  0.45  &  0.49 \\
$8_{26}$--$9_{19}$   &  159.485  &  1324.0  &  0.76  &  0.25  &  0.70  &  0.01  &  0.14  &  0.09 \\
$5_{32}$--$5_{23}$   &  160.510  &  642.4  &  18.7  &  5.5  &  2.6  &  0.09  &  1.2  &  1.6 \\
$7_{34}$--$7_{25}$   &  166.815  &  1125.7  &  5.2  &  0.58  &  0  &  0.06  &  0.33  &  0.17 \\
$6_{24}$--$6_{15}$   &  167.035  &  781.1  &  10.9  &  3.7  &  0.80  &  0.26  &  1.0  &  0.82 \\
\hline
LW5   &    &    &    &    &    &    &    &  \\
$7_{35}$--$6_{42}$   &  169.739  &  1090.3  &  15.5  &  0.19  &  0.46  &  0  &  1.4  &  4.1\\
$6_{33}$--$6_{24}$   &  170.139  &  867.3  &  35.4  &  9.9  &  2.2  &  0.55  &  2.1  &  1.4\\
$5_{33}$--$6_{06}$   &  174.607  &  642.7  &  48.8  &  14.4  &  4.3  &  4.9  &  2.6  &  6.7\\
$3_{03}$--$2_{12}$   &  174.626  &  114.4  &  3.2  &  1.4  &  0.34  &  0.25  &  0.10  &  0.71\\
$4_{32}$--$5_{05}$   &  174.920  &  468.1  &  5.7  &  0.10  &  1.6  &  0.63  &  2.3  &  0.60\\
$2_{12}$--$1_{01}$   &  179.527  &  34.2  &  52.7  &  11.2  &  5.0  &  5.5  &  0  &  8.8\\
$2_{21}$--$2_{12}$   &  180.488  &  114.4  &  26.4  &  8.5  &  2.1  &  0.26  &  3.7  &  4.2\\
$4_{13}$--$4_{04}$   &  187.111  &  319.5  &  15.1  &  4.9  &  0  &  0  &  0.84  &  1.3\\
$8_{54}$--$9_{27}$   &  187.810  &  1729.3  &  3.7  &  0.91  &  0  &  1.8  &  0.43  &  3.6\\
$6_{43}$--$7_{16}$   &  190.437  &  1013.2  &  6.8  &  0.64  &  1.4  &  0  &  2.3  &  1.3\\
$6_{33}$--$5_{42}$   &  194.422  &  877.8  &  14.6  &  4.4  &  1.5  &  4.8  &  0.88  &  7.9 \\
\hline
\end{tabular}
\end{center}
\end{table*}

\end{document}